# A Chemical Route to Graphene for Electronics and Spintronics Device Applications

Santanu Sarkar*[$]

[$]Center for Nanoscale Science and Engineering, Departments of Chemistry and Chemical & Environmental Engineering, University of California, Riverside, CA 92521-0403, United States
*E-mail: ssark002@ucr.edu

[$]Present address: Intel Corporation, Logic Technology Development Group, Ronler Acres Campus, Hillsboro, OR 97124, United States
*E-mail: santanu.sarkar@intel.com

ABSTRACT: The development of selective high precision chemical functionalization strategies for device fabrication, in conjunction with associated techniques for patterning graphene wafers with atomic accuracy would provide the necessary basis for a post-CMOS manufacturing technology. This requires a thorough understanding of the principles governing the reactivity and patterning of graphene at the sub-nanometer length scale. This article reviews our quest to delineate the principles of graphene chemistry – that is, the chemistry at the Dirac point and beyond, and the effect of covalent chemistry on the electronic structure, electrical transport and magnetic properties of this low-dimensional material in order to enable the scalable production of graphene-based devices for low- and high-end technology applications.

*The new carbon age,[1] which is the third wave in the carbon revolution, has witnessed overwhelming interest in low-dimensional carbon materials, with particular attention to graphene, the newest member of the series of carbon allotropes. This two-dimensional form of pure $sp^2$ hybridized carbon – the giant molecule[2]* of *atomic thickness has garnered tremendous attention among both physicists and chemists and has provided a test-bed for fundamental and device physics,[3,4] and a unique chemical substrate.[5-9] In line with theoretical predictions, charge carriers in graphene behave like massless Dirac fermions, which is a direct consequence of the linear energy dispersion relation.[10] Such features serve to recommend graphene*

*for mechanical, thermal, electronic, magnetic and optical applications, but the absence of a band-gap in graphene makes it unsuitable for conventional field effect transistors (FETs),[11,12] and its lack of solution processability remains to be resolved.[13] These issues are potentially amenable to solution by chemical techniques, but the effect of chemistry on the mobility of functionalized graphene devices is an imposing challenge.[14]*

Dimensionality defines the physical and chemical behavior of a material and distinguishes one material from the other even among those of the same chemical composition;[15] while the chemical concepts of structure and hybridization lead to the same conclusion.[16-18] From the standpoint of both physics and chemistry, the two-dimensional (2D) graphene materials with atomically flat surface are remarkably different from that of the quasi-zero-dimensional (0D) fullerenes, and the one-dimensional (1D) carbon nanotube materials. The experimental isolation of graphene atomic crystals by Geim and Novoselov at the University of Manchester and the realization of epitaxial graphene devices by de Heer and co-workers at Georgia Tech, triggered research interest in this 2D material in 2004.[3,19] As a chemical substrate graphene is unique – any two adjacent carbon atoms in graphene are crystallographically non-equivalent (and belong to the A and B sub-lattices), but chemically they are equivalent.[11,20]

The pursuit of the chemical functionalization of graphene is based on a number of motivations: (i) definition of 1-D ballistic wires for interconnects, (ii) band-gap engineering for transistors (modification of the electronic structure of graphene),[8] (iii) creation of magnetism in graphene for applications in spintronics,[21,22] (iv) fabrication of dielectrics, (v) bulk preparation of solution-processable derivatives for a broad range of applications including hydrogen storage, thermal interface materials (TIMs), nano-bio hybrid composites,[23] and (vi) as the solid state counterpart of classical small molecule electrocyclic organic reactions, including

Diels-Alder chemistry,[12,20] and the Claisen rearrangement.[24] The availability of solution processable graphene can make an important contribution to emerging fields such as printable electronics, and is expected to enable chemical modification, purification, and the transfer of graphene from the solution phase to substrates by means of spin-, spray-, drop-, or dip- casting methods.[25]

However, the basal plane chemical modification of graphene is not straightforward because normal aromatic substitution reactions cannot be applied, and in this respect graphene chemistry resembles that of the fullerenes and carbon nanotubes, but without the role of strain in promoting addition chemistry.[17,18] Nevertheless, it has already been demonstrated that the atomically flat surface of graphene provides an opportunity to apply carbon-carbon bond formation chemistry to graphene with subsequent creation of $sp^3$ carbon centers in place of $sp^2$ carbons in the honeycomb lattice of graphene.[9,22] This chemistry has a pronounced effects on the electronic and phonon properties of graphene as is evidenced by obvious changes in the Raman spectra, transport properties, magnetic behavior and scanning tunneling microscopic images in the resulting products. The covalent modification of the two-dimensional π-electron system of graphene provides a novel protocol to impart patterning that can modulate the energy band gap, influence electron scattering, affect the flow of current by creating dielectric regions over the graphene wafer,[22] and potentially address some of the issues in the fabrication molecular level electronic circuitry.[26] The fundamental concept is the generation of new carbon-carbon bonds to redirect the electronic conjugation pathway and to form specific conducting, ballistically conducting, insulating, semiconducting, and magnetic patterns.

In this article we focus on two graphene chemistries: (i) nitrophenyl (NP) radical addition to graphene, which leads to formation of one C−C bond, and therefore one $sp^3$ carbon center (together with a free spin) in the graphene lattice per added functionality,[8,27] and (ii) Diels-Alder (DA) chemistry, which leads to the simultaneous creation of two C−C bonds, and therefore a pair of $sp^3$ carbon centers in the graphene lattice (which are necessarily in the A- and B-sub-lattices, and do not create a free spin) per added functionality.[12] The ensuing discussion of the NP and DA functionalization of graphene is aimed at providing a simple proof-of-concept for the application of covalent chemistry in graphene for electronics, photonics, and spintronics and in order to give an update of progress in the field.

**Microscopic and spectroscopic detection of graphene flakes**

Despite the atomic thickness of graphene, optical microscopy can be conveniently employed to identify a monolayer of graphite (single-layer graphene, SLG, with number of layers n = 1), bi-layer graphene (BLG, n =2), tri-layer graphene (TLG, n = 3), few-layer graphene (FLG, n ≥ 4) along with thicker graphite flakes [n = $\infty$, such as highly oriented pyrolytic graphite (HOPG)] from the color contrast (phase contrast) when the graphene flakes rest on the top of an oxidized Si wafer (Fig. 1).[15] SLG on an $SiO_2$ substrate, although atomically thin, has the capability to interfere with the optical path of reflected light, and consequently results in a change in interference color with respect to bare $Si/SiO_2$ substrate (typically about 300 nm $SiO_2$, and is purple-to-violet in color).[15] As can be seen in Fig. 1, SLG is very light violet in color (Fig. 1a), BLG appears as violet (Fig. 1b), TLG as dark violet (Fig 1c), while FLG flakes are blue in color (Fig. 1b). Microscopic quality and macroscopic continuity are two essential ingredients in judging the quality of a graphene sample. In contrast to exfoliated graphene, epitaxial graphene (EG) samples grown by vacuum graphitization of SiC(0001) are almost transparent on SiC (Fig. 1d). In the case of supported graphene, the substrate has a strong influence on the subsequent chemistry and device performance. On the other hand, the chemistry of epitaxial graphene (EG) on SiC substrates shows the effects of the interface layer between the graphene monolayers and the underlying SiC substrate.[8,28,29]

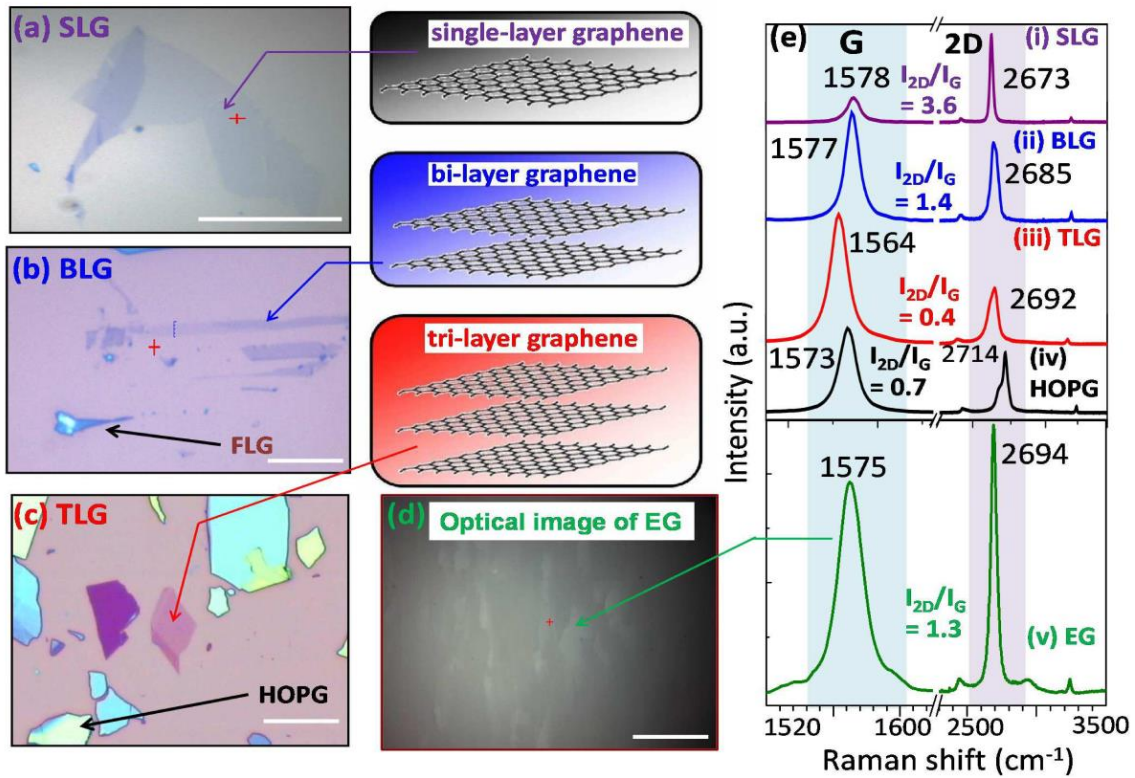

*Fig. 1* Optical microscopic image of (a) single-layer graphene (SLG), (b) bilayer grahene (BLG), (c) trilayer graphene (TLG), along with few-layer graphene (FLG in b), and graphite (HOPG in c), obtained by micromechanical cleavage of graphite and imaged on an oxidized Si wafer. The corresponding chemical structures are shown in the right frame. (d) Optical image of epitaxial graphene (EG) grown by vacuum graphitization on SiC(0001). Scale bar is 20 μm. (e) Raman spectral signatures ($\lambda_{ex}$ = 532 nm) of the corresponding (i) SLG, (ii) BLG, (iii) TLG, (iv) HOPG, and EG (after subtraction of Raman signals due to SiC).

Beyond microscopic visualization, Raman spectroscopy provides the most convenient and powerful approach for the detection of the graphene flakes. In the Raman spectra of graphene, the G peak (frequency, $\omega_G$ ~ 1580 cm$^{-1}$), arises from a first order Raman effect where the energy of the scattered incident monochromatic light is proportional to the energy of quantized lattice vibrations ($E_{2g}$ phonon) created by the scattering process.[22,30-33] On the other hand, the 2D band ($\omega_{2D}$ ~2670 cm$^{-1}$, also referred to as the G' peak) is a second order Raman effect, which arises from lattice vibrations when first order processes activate another phonon. In the case of a single-layer graphene (SLG), the 2D peak appears as a single peak and the intensity of 2D peak is generally higher than the intensity of G-peak [$I_{2D}/I_G \geq 1$, Fig. 1e(i)].

As discussed later, the covalent chemical modification of graphene, which is usually accompanied by conversion of $sp^2$ hybridized carbons to $sp^3$, leads to the activation of the $A_{1g}$ breathing vibration mode and this results in the appearance of a sharp D-peak ($\omega_D$ ~1345 cm$^{-1}$); broad D-peaks can also be seen in physically defective graphitic materials, such as graphene nanoribbons (GNRs), the edges of graphene, disordered graphene samples, and in graphene oxide.[34]

Raman spectroscopy provides a wealth of information about the number of graphene layers (based on the position and shape of the 2D band, and the ratio of the intensities of the 2D to G band, Fig. 1e),[30] quality of the samples,[35] types and degree of doping in the sample (based on observed shift of G and 2D band),[36] and can even provide insight into the mobility of the graphene devices.[37]

**Reactivity of graphene**
Based on the observed chemical behavior of graphene, a number of structural and electronic features have been found important in understanding the reactivity of graphene as a chemical substrate. (i) **Role of dangling bonds**: Edges containing dangling bonds are the most reactive,[38] and within basal plane chemistry, the thermodynamically (energetically) favorable processes involve the pairwise chemisorption of functional groups in different sublattices, rather than on the same sublattice.[6,39] Theoretical calculations suggest that the pairwise chemisorption of a species in different sublattices is favored by 0.5 eV per addition.[39,40] (ii) **Minimization of geometric strain**: In analogy with the fullerenes and carbon nanotubes, which contain curved graphitic surfaces,[17,18] geometrically strained areas and ripples in graphene undergo preferential reactivity in order for these regions to relax by rehybridization.[6,34,41] Strain engineering on the surface lattice of graphene in a periodic manner can control the reactivity and degree of functionalization of graphene.[6] (iii) **Role of aromatic** ***sextets in graphene rings at basal plane and edges***: The Clar sextet is the most stable resonance structure and those graphene structures that maximize the number of Clar sextets will be preferred. At the graphene edges, which can be either zig-zag or arm-chair structures, the attainment of aromatic sextets is frustrated in most of the rings where zig-zag edges are concerned, and are therefore thermodynamically unstable and more reactive than arm-chair edges.[6,42,43] (iv) ***Chemistry at the Dirac Point – frontier molecular orbitals and conservation of orbital symmetry***: The graphene valence band (HOMO) and conduction band (LUMO) cross at the Dirac point, which defines the work function (W = 4.6 eV). Consequently, the HOMO and LUMO of graphene form a degenerate pair of orbitals at this point in momentum space with the same ionization potential (IP) and electron affinity (EA), and these states determine the reactivity. Pericyclic reactions are subject to the Woodward-Hoffmann rules, and inspection of the orbital symmetries of the degenerate pair of half-occupied HOMO and LUMO band orbitals at the Dirac point confirms that with the appropriate orbital occupancies, both diene and dienophile reaction partners should undergo Woodward-Hoffmann allowed, concerted Diels-Alder reactions with graphene.[12,20] Because of the orbital crossing at the Dirac point, the $\pi$-bonds in graphene can access diene or olefinic (quinonoid) resonance structures.[12,20] This behavior is manifested by the reactivity of graphene with electron-rich dienes in Diels-Alder chemistry (as diene and dienophile),[12,20] in nitrene addition chemistry,[44] in Bingel [2+1] cyclopropanation reaction,[45] and 1,3-dipolar cycloaddition reactions.[46]

**Distinction between covalent chemistry and ionic chemistry when applied to single-walled carbon nanotubes**

In the present article we illustrate the effects of covalent chemistry on the electronic and magnetic properties of graphene. The distinction between covalent chemistry and ionic doping chemistry is clearly illustrated by the effect of the covalent addition of dichlorocarbene species to single-walled carbon nanotubes (SWNTs), which leads to formation of a cyclopropane ring and the generation of a pair of sp$^3$ carbon centers in SWNTs (see inset in *Fig. 2a*). Progressive addition of dichlorocarbene to SWNTs[47,48] leads to a gradual decrease in conjugation length of the periodic extended conjugation of the metallic SWNTs, resulting in removal of states at the Fermi level and a concomitant decrease in the $M_{00}$ band intensities in the far-IR optical absorption spectra (*Fig. 2a*). In contrast to the covalent carbon−carbon bond formation chemistry, ionic doping of SWNTs by molecular bromine ($Br_2$) introduces holes into the valence band of the semiconducting SWNTs, which increases the concentration of free carriers and the spectral weight of the transitions at the Fermi level in the far-IR part of the spectrum (*Fig. 2b*). Parallel insight into the application of similar covalent chemistry in the band-gap engineering of graphene can be obtained from discussions of radical addition chemistry and Diels-Alder chemistry, which is the subject of the present article.

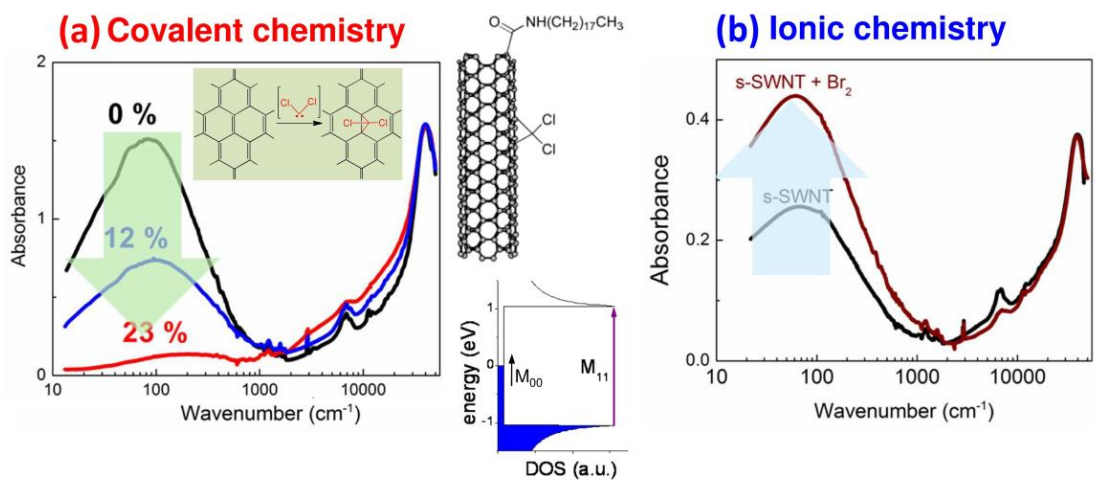

*Fig. 2* Distinction between the effects of covalent chemistry and ionic chemistry on the energy spectrum of single-walled carbon nanotube (SWNTs) materials. Changes in the optical absorption spectra of SWNTs due to (a) progressive addition of dichlorocarbene species to the sidewalls of SWNTs, which gradually suppresses the absorption band in the far-IR region (a characteristics of the metallic SWNTs transitions, $M_{00}$), and (b) ionic doping chemistry after addition of molecular bromine ($Br_2$) SWNTs (adapted from ref [47]).

**Radical addition chemistry applied to graphene**

The addition of aryl radicals to graphene[27,38,49-52] leads to the creation of carbon−carbon (C−C) bonds between graphene and aryl moieties, and consequently is accompanied by the formation of $sp^3$ centers in place of $sp^2$ hybridized carbons in the graphene honeycomb lattice (Fig. 3a). One of the most popular approaches for the generation of aryl radicals involves the spontaneous reduction of the corresponding diazonium salts.[27] The spontaneous reduction of *p*-nitrophenyl diazonium tetrafluoroborate salts on graphene surfaces (EG,[27,33,50] exfoliated SLG, and BLG)[14,33], leads to *p*-nitrophenyl (NP) functionalized graphene derivatives (*Fig. 3a,3b*). The application of this chemistry in EG samples brings about pronounced changes in the Raman spectra of NP-EG [$I_D/I_G$ =2.4, compared with $I_D/I_G$ = ~0 in pristine EG] (*Fig. 3c*) as well as the AFM height images (EG in *Fig. 3d* and NP-EG in *Fig. 3e*).

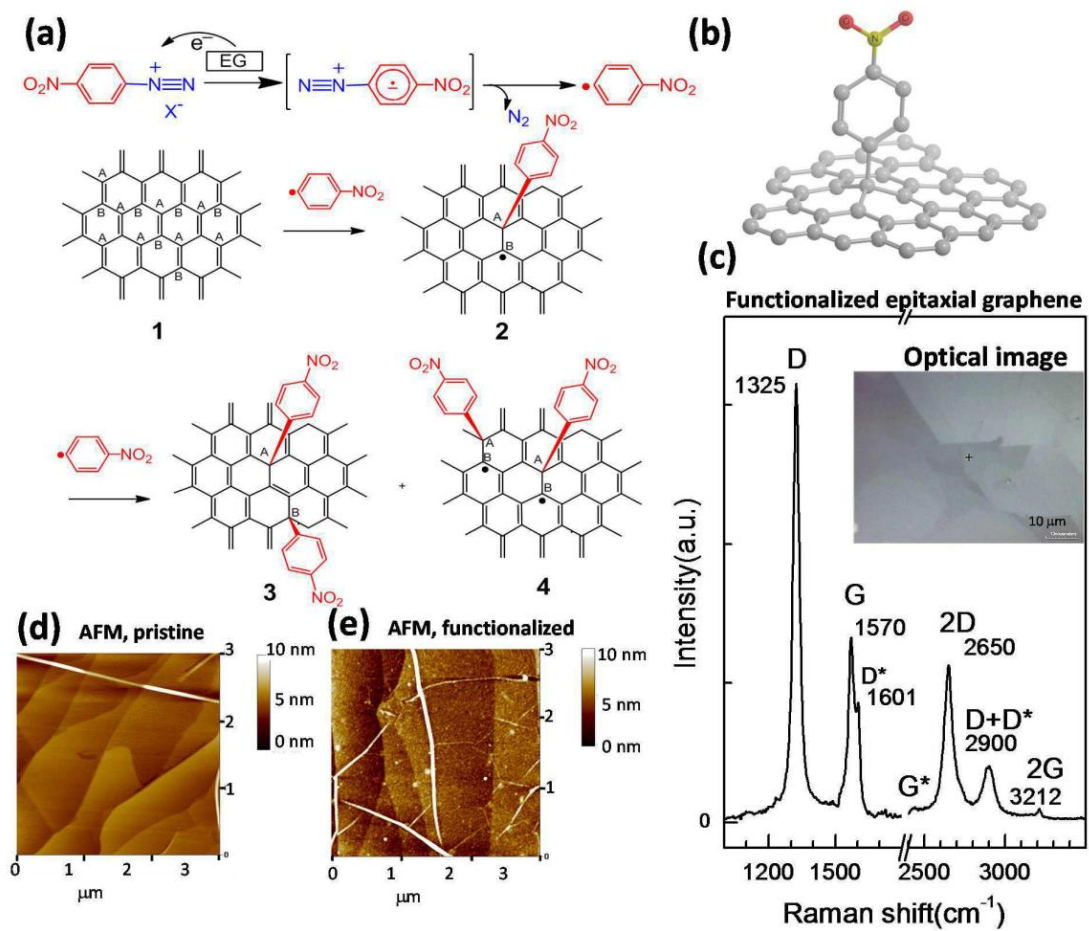

*Fig. 3* Reactivity of graphene in spontaneous electron transfer chemistry. (a) Schematics of spontaneous reduction of *p*-nitrophenyl (NP) diazonium salts on graphene surfaces **1** leading to covalent attachment of NP groups to graphene. The addition of the first radical leads to the product **2**, which is expected to be paramagnetic, while the addition of a second NP radical may lead to either product **3** (diamagnetic), and (or) product **4** (biradical).[22] (b) Representation of a three-dimensional image of NP-EG. (c) Raman spectra ($\lambda_{ex}$ = 532 nm) after functionalization of EG with NP groups (NP-EG) shows a pronounced increase of the D-band intensity, whereas pristine epitaxial graphene (EG) shows virtually no D-band. Comparison of AFM images of (d) pristine EG and (e) NP-EG. Adapted with permission from ref [33], © 2010 American Chemical Society.

## Chemical reactivity as a function of the substrate and number of graphene layers

Graphene is available in a variety of forms, and there is already strong evidence that the chemical and physical properties of the material are sensitive to the substrate and the number and orientation of the graphene sheets.[8,20,53] Raman

-10-

spectroscopy can be conveniently employed to track the progress of reactions, and to estimate the differences in the reactivity between SLG, BLG, FLG, EG and HOPG. The SLG sheet is reported to be almost 10 times more reactive than BLG graphene in its reactivity towards *p*-nitrophenyl diazonium tetrafluoroborate, and the graphene edges are reported to be at least two times more reactive than the bulk SLG sheets.[38] A number of authors have focused on adsorbed intermediate in NP addition chemistry,[51] and in some cases the reaction leads to graphene doping.[54] The spontaneous reaction of SLG (*Fig. 4a*) with *p*-nitrophenyl diazonium tetrafluoroborate in acetonitrile under argon leads to a pronounced D band ($\omega_D$ ~1336 cm$^{-1}$, with $I_D/I_G$ = 2.6 , *Fig. 4c*), while the same chemistry on BLG (*Fig. 4b*) leads to a very weak D-band (with $I_D/I_G$ = 0.4, *Fig. 4d* ) and the corresponding AFM height images of pristine SLG and BLG on an oxidized Si wafer is shown in *Fig. 4f*, while *Fig. 4g* shows the increased roughness in AFM height image of NP-SLG.[27,33].

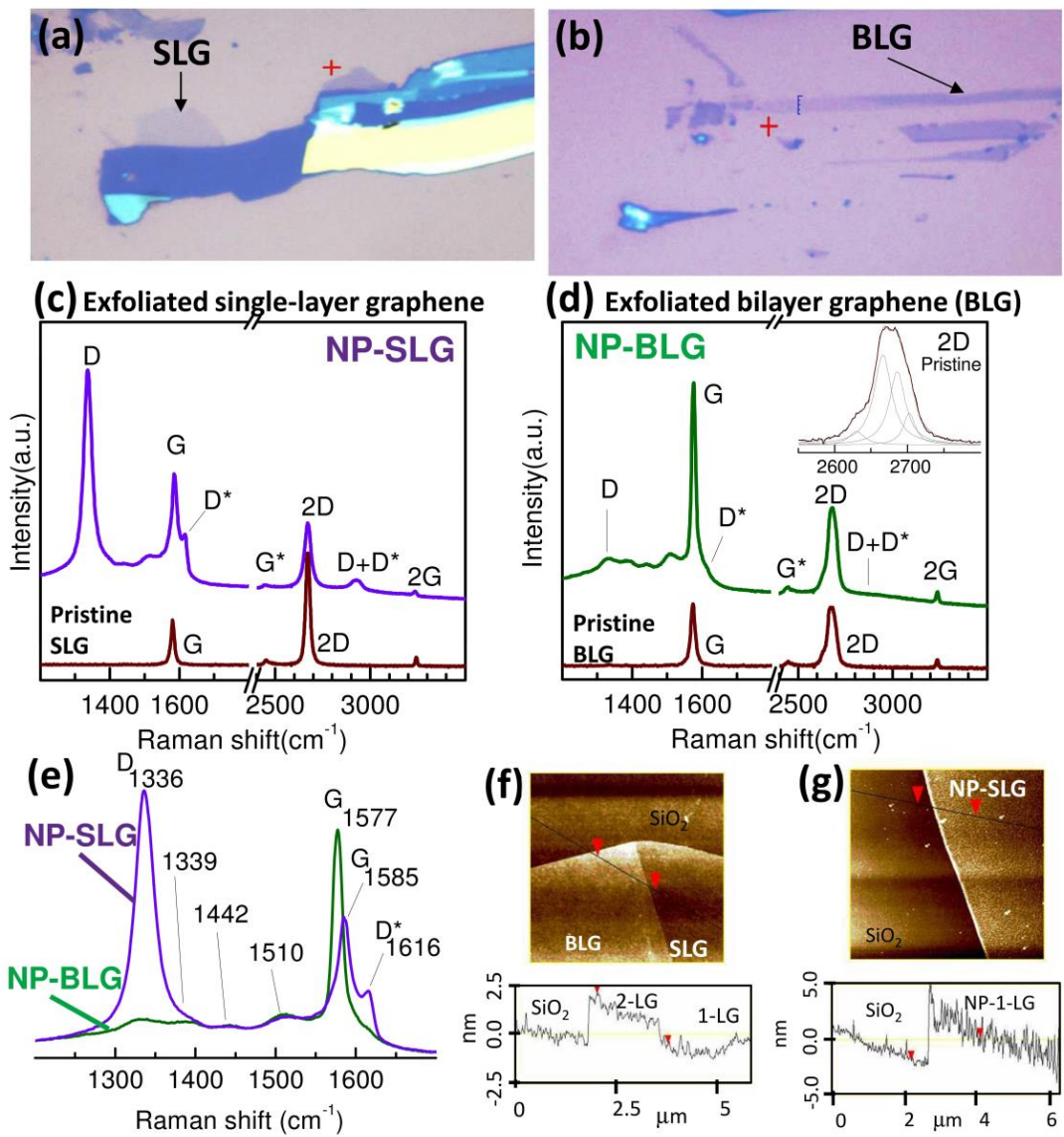

*Fig. 4* Changes in chemical reactivity with the number of graphene layers. Optical images of (a) SLG, and (b) BLG flakes on an oxidized Si wafer. Raman spectra ($\lambda_{ex}$ = 532 nm) after nitrophenyl radical addition to (c) SLG, and (d) BLG flakes, whereas the comparison of Raman spectra of NP-SLG and NP-BLG is shown in (e). The SLG flakes show higher reactivity in radical addition chemistry than BLG as is seen by evolution of D-band in the Raman spectrum. AFM height images of (f) pristine SLG, and BLG on the same substrate (Si/SiO$_2$), and (g) NP addition product to SLG (NP-SLG), which shows increased surface roughness after functionalization. Adapted with permission from ref [33], © 2010 American Chemical Society.

**Room-temperature ferromagnetism in NP-EG: quasi-localized π-radicals**

Addition of a single hydrogen atom or a single carbon σ-radical to graphene leads to the formation of a delocalized spin in the graphene π-system (structure **2** in *Fig. 3a*, where addition has occurred in the A sublattice). If the second radical adds in a different sublattice (B, as in the structure **3**, in *Fig 3a*) a spin-paired diamagnetic structure results, whereas addition to the same sublattice (A, as in structure **4** in Fig. 3a) gives an open-shell structure. The parallel alignment of the spins in a single sublattice to give ferromagnetism (B of **4**, in *Fig. 3a*) is favored because the spins are confined to the same sublattice,[55-58] and thus Hund's rule favors triplet coupling in species such as **4** because of the Pauli exclusion principle, which forbids electrons with the same spin from occupying the same space and thus serves to reduce the Coulomb repulsion energy in this state in comparison with the spin-paired singlet.[59] Magnetic measurements show a small magnetization in pristine EG samples at all temperatures, which is attributed to either defects or impurities in the epitaxial graphene on SiC crystals.[21] The difference in the M-H data between NP-EG and pristine EG reflects the effect of the NP functionalization of the top layer, and it shows a nonlinear dependence on a magnetic field strength and a clear hysteresis.[21,22] Both the coercivity ($H_c$) and the saturation magnetization ($M_{sat}$) vary from sample to sample,[21] and it will be important to establish reproducible behavior if the results are to find applications.[60] The data indicates that the magnetism is confined to random clusters on the surface of the EG with varying values of the coercive field and is indicative of a distribution in the sizes and interaction strengths of the magnetic regions, which is typical of magnetic objects that are close to the border of ferromagnetism (ferrimagnetism) and superparamagnetism.[8,21,22]

**Radical addition chemistry and the introduction of an energy gap**

The zero-band-gap semiconducting nature of graphene limits the applicability of graphene in conventional FET devices. However, covalent chemistry carried out on epitaxial graphene (*Fig. 5a-c*)[8,22,27,33,50] and exfoliated SLG (suspended film, *Fig. 5d-g*)[14] by use of simple solution chemistry techniques suggests the applicability of this technique for band-gap engineering of graphene devices. Theoretical calculations on fully hydrogenated graphene [graphane, $(CH)_n$], which requires the conversion of all $sp^2$ carbons of graphene to $C(sp^3)$–H bonds, indicates band gaps of 3 eV,[40,61] and 5.4 eV.[62] Similarly, the widely-studied stoichiometric fluorinated derivative of graphene (fluorographene), which is a thermally stable alternative to graphane, is an insulator with an optical gap close to 3 eV.[63,64] Steric considerations and the single-sided functionalization process of non-suspended SLG preclude high coverage with NP groups, and even the 25% coverage model shown in *Fig. 5a*, is not attainable; and measurements indicate a coverage of 10-20%.[8,65]

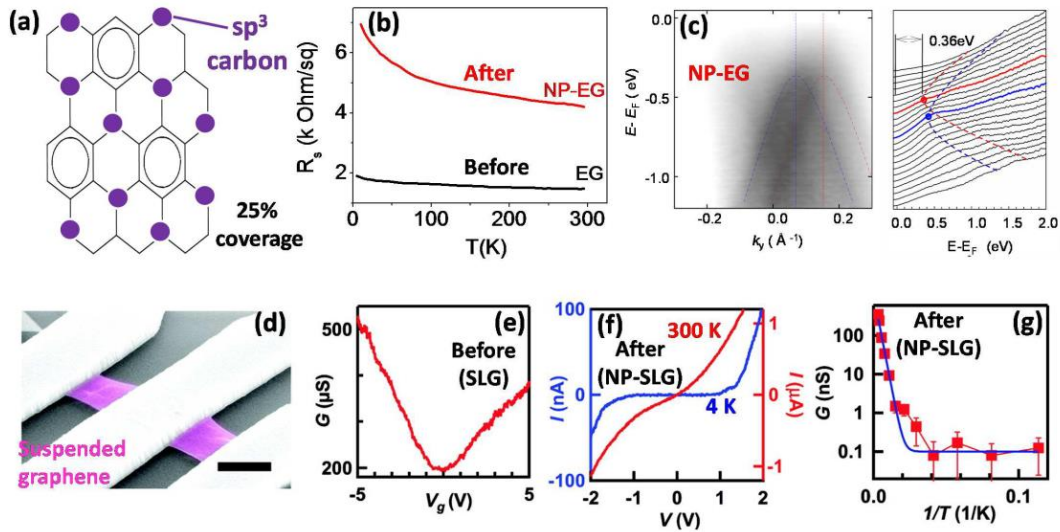

*Fig. 5* Effect of covalent chemistry on transport properties of graphene. (a) Schematics of a model radical addition process on the graphene lattice, which gives 25% coverage.[8] (b) Change in resistance and its temperature dependence after NP functionalization of EG.[27,50] Reprinted with permission from ref[27], © 2009 American Chemical Society. (c) Angle-resolved photoelectron emission spectroscopy (ARPES) of NP functionalized EG (NP-EG) showing two diffuse bands highlighted by the dashed lines, corresponding to Dirac cones with a band edge 0.36 eV below the Fermi level.[22,33] Adapted with permission from ref[33], © 2010 American Chemical Society. (d) False-color SEM image of a suspended graphene device. (e) $G(V_g)$ of a typical suspended device before functionalization (pristine EG). Scale bar: 1 μm. (f) I–V curves of a suspended functionalized device (NP-EG) at $V_g = 0$ and T = 300 K (red curve, right axis) and 4 K (blue curve, left axis), respectively. (g) Linear response $G$ vs $1/T$. The solid line is the best-fit to $G(T) = G_0 + A \exp(-E_A/k_B T)$, where $E_A \sim 40$ meV. Reprinted with permission from ref[14], © 2011 American Chemical Society.

The NP radical addition chemistry has a pronounced effect on the transport properties of graphene and the resistance of pristine EG and NP-EG as a function of temperature is shown in *Fig. 5b*.[27] The pristine EG (5-7 graphene layers) shows ideal semimetallic behavior with zero or small energy gap; the increase in resistance with decreasing temperature is attributed to the decreasing carrier density as has been previously reported for sub-10 nm thick graphite samples.[66-68] The NP functionalization of EG results in an increase in the room-temperature resistance from 1.5 to 4.2 kΩ/square, and a more pronounced temperature dependence; the semiconducting nature of the NP-EG is supported by the observation of a band gap of 0.36 eV in angle-resolved photoelectron emission spectroscopy (ARPES) measurements.[33] This study suggests that surface covalent functionalization of the top layer of epitaxial graphene is capable of influencing the bulk properties of the EG sample.

The ARPES measurements in Fig. 5c show the modified band structure of graphene at the K point;[33] in NP- EG the linear bands of EG are transformed into massive bands shifted ~0.36 eV below the Fermi level and constant energy cuts (Fig. 5c, right-hand figure) show that an energy-gap has opened in NP-EG.[8]

In contrast to the EG substrates (EG/SiC), which allow only one-sided functionalization of the topmost graphene layer, the suspended graphene (SG) films (*Fig. 5d*) provide the opportunity for double-sided covalent functionalization which is shown to produce a granular metal at low NP coverage, and a gapped semiconductor at high NP coverage.[14] The pristine free-standing graphene (SG) membranes typically show a mobility of 5,000-15,000 cm$^2$ V$^{-1}$s$^{-1}$ (*Fig. 5e*).[14] After NP functionalization of SG, the mobility decreased to 50–200 cm$^2$V$^{-1}$s$^{-1}$, the I–V curves are seen to be non-linear even at 300 K, and at 4 K the conductance is effectively zero (*Fig. 5f*). In *Fig. 5g* it can be seen that the zero-bias conductance [G($V_g$ = 0 V)] decreases exponentially with 1/T at high temperature and crosses over to a constant value for T < 30 K. The data in *Fig. 5g* can be fit to the equation: $G(T) = G_0 + A \exp(-E_A/k_B T)$, where the activation energy, $E_A$ ~ 39 ± 10 meV, and $G_0$ is the background conductance; thus for double-sided NP addition to SG the energy gap is estimated as $2E_A$ ~ 80 meV at room temperature.[14] Thus the surface density of the covalently linked functional groups is able to change graphene from a gapless semimetal, to a granular metal, which displays variable range hopping with a low temperature localization-induced gap, to a semiconductor with a transport gap.[14]

**Chemistry at the Dirac Point: Diels-Alder chemistry of graphene**

This reaction was first documented by Otto Diels and his student Kurt Alder in 1928,[69,70] for which they were awarded the Nobel Prize in Chemistry in 1950. The prototypical Diels-Alder [4+2] cycloaddition process involves the reaction between a diene (generally a 4π–electron system, such as 1,3-butadiene) and a dienophile (generally a 2π–electron system, such as ethylene) leading to the formation of a cyclohexene ring system (*Fig. 6a*). These [4+2] cycloaddition processes are now known as pericyclic reactions and the recognition of the importance of orbital symmetry and the frontier molecular orbitals (FMOs) in such processes led to the award of the 1981 Nobel Prize in Chemistry to Kenichi Fukui and Roald Hoffmann.

The zero-band-gap electronic structure of graphene (*Fig. 6b, 6c*) arises from the crossing of the valence and conduction bands at the Dirac point (K), which dictates that the highest occupied molecular orbital (HOMO) and lowest unoccupied molecular orbital (LUMO) energies are coincident with the work function, and thus graphene has an exceptionally low ionization potential (HOMO) and an the exceptionally high electron affinity (LUMO)(*Fig. 6d*).[12,20] These factors, together with the orbital symmetries of the degenerate HOMO and LUMO bands at the Dirac point, enable graphene to participate in the Diels-Alder reaction and to show dual reactivity - that is, graphene behaves as either diene or dienophile (*Fig. 6e, 6f*) when paired with the appropriate reaction partner.[20] To elaborate, graphene can function as either diene when paired with tetracyanoethylene (TCNE) and maleic anhydride (MA) or as dienophile when paired with 2,3-dimethoxybutadiene (DMBD) and 9-methylanthracene (9-MeA).[12]

The DA chemistry leads to the simultaneous transformation of a pair of $sp^2$ carbons in the graphene honeycomb lattice to either 1,2– or 1,4–$sp^3$ carbon centers, introducing a barrier to electron flow by opening a band gap and allowing the generation of insulating and semiconducting regions in graphene wafers.

The Diels-Alder reaction is one of the most powerful and elegant reactions in synthetic organic chemistry, and the reaction encompasses a number of features, which are important in the covalent modification of graphene: (i) catalysts are not required, (ii) efficiency under mild reaction conditions, (iii) simultaneous formation of a pair of $sp^3$ carbons (spin-paired Kekule structures), (iv) exclusion of the possibility of generating conjugated $\pi$-radicals, (v) reactive towards graphene edges leading to elongation and quenching of graphene edges, (vi) absence of by-products, and (vii) availability of simple thermal retro-DA reactions, which allow the regeneration of the starting materials.[8,12]

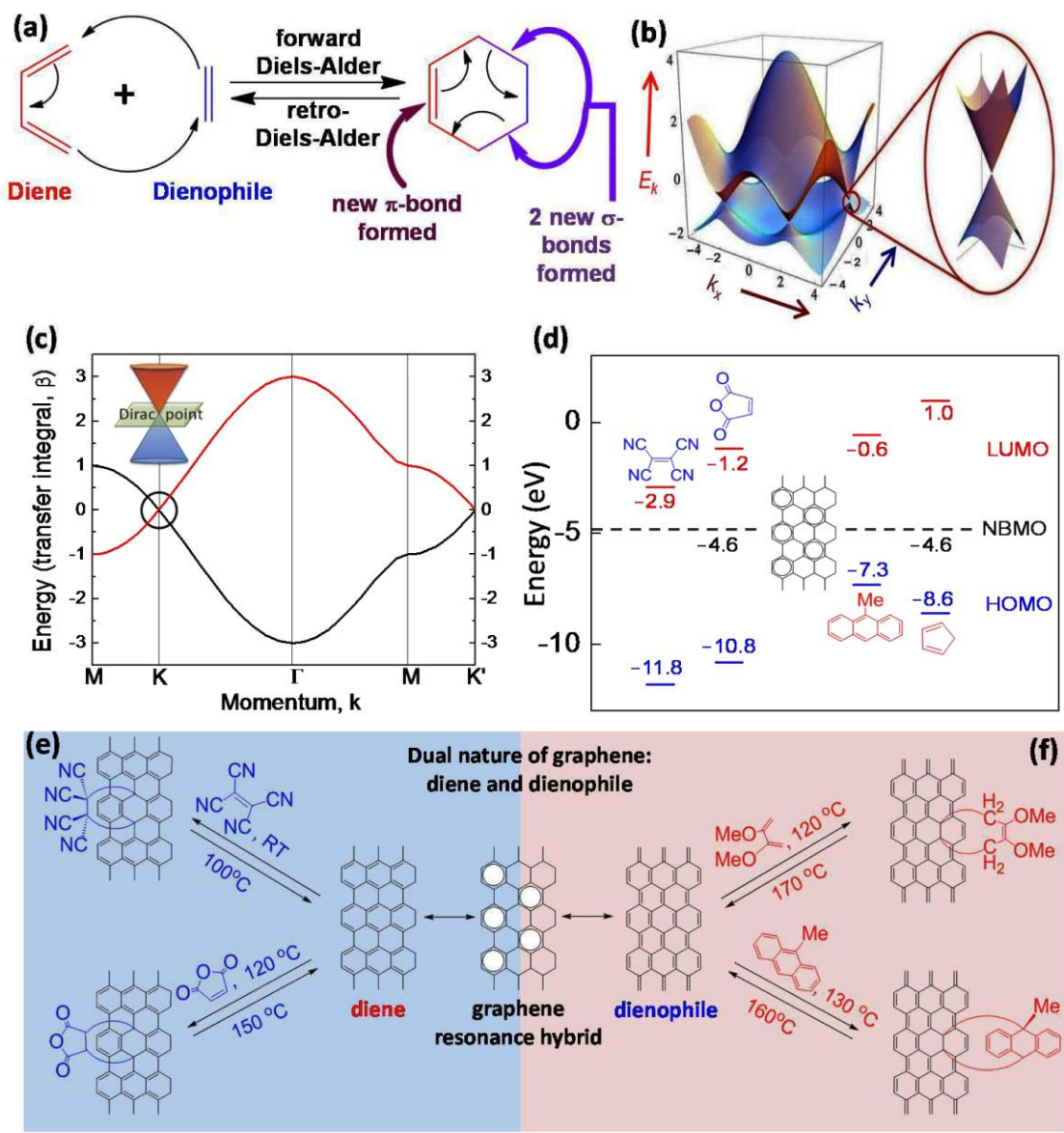

*Fig. 6* (a) Schematic illustration of the Diels-Alder [4+2] cycloaddition between 1,3-butadiene (diene) and ethylene (dienophile). (b) Electronic band dispersion in the graphene honeycomb lattice. Left: energy spectrum (in units of $t$), with $t = 2.7$ eV and $t' = -0.2t$. Right: expanded view of energy bands close to one of the Dirac points.[11] (c) Dispersion of the graphene energy band in momentum space within simple tight-binding (HMO) theory as a function of the resonance or transfer integral ($\beta$, t ~ 3 eV).[20] (d) Orbital energies of selected dienes and dienophiles as obtained from ionization potentials (HOMO, −IP), electron affinities (LUMO, −EA), and the work function of graphene (W = (−)4.6 eV). The neutrality point in graphene corresponds to the energy of the carbon-based nonbonding molecular orbital (NBMO).[20] Schematic illustration of the Diels-Alder chemistry of graphene as (e) diene, and (f) dienophile.[8,20] Adapted with permission from ref[20], © 2012 American Chemical Society.

The chemical behaviour of graphene as a diene is illustrated in *Fig. 7* by its reactivity with the electron-withdrawing dienophile, tetracyanoethylene (TCNE). Raman spectroscopy is employed to monitor the progress of the reaction and to track the differential reactivity of SLG, FLG, and HOPG; *Fig. 7c* shows an increase of $I_D/I_G$ ratio in the TCNE-SLG adduct to 2.53 from 0.03 in pristine SLG, while the $I_D/I_G$ ratio is 0.28 in TCNE-FLG and 0.17 in TCNE-HOPG for reactions conducted under identical conditions.[12,20] The differential evolution of the D-band in the presence of the same DA chemistry suggests the following order of reactivity in DA chemistry: SLG >> FLG > HOPG.

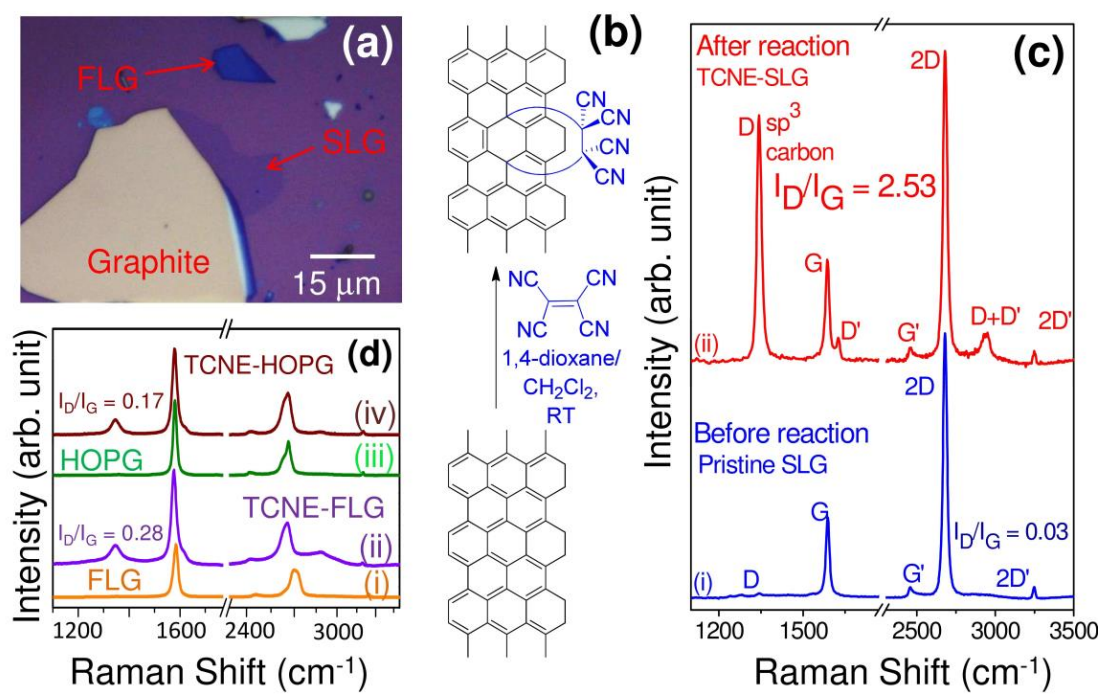

*Fig. 7* Graphene as diene: (a) optical micrograph of single-layer (SLG), few-layer graphene (FLG) and graphite (HOPG). Contrast in the image is increased by 30% to enhance clarity. (b) Schematics of the room-temperature reaction of graphene (as diene) with tetracyanoethylene (TCNE, dienophile). Differential reactivity of (c) SLG, (d) FLG, and graphite (HOPG) in the Diels-Alder chemistry with TCNE is manifested by the evolution of the Raman D-band.

Similarly, the chemical behaviour of graphene as a dienophile is illustrated in *Fig. 8* by its reactivity with electron-rich dienes, such as 2,3-dimethoxy-1,3-butadiene (DMBD), and 9-methylanthracene (9-MeA). The DMBD-EG adduct (*Fig. 8a*) shows an increase of $I_D/I_G$ ratio (from 0 to 0.5) after functionalization, and the room-temperature resistance increases by 60%, and shows an activated, non-metallic temperature-dependence over the whole temperature range (*Fig. 8c*). In the functionalization of EG with 9-MeA (*Fig. 8d*), the Raman spectra show that the pristine EG (*Fig. 8e*) is almost defect free (absence of D-band), while DA functionalization leads to a moderate to high D-peak intensity over a large fraction of the graphene wafer (*Fig. 8f*).

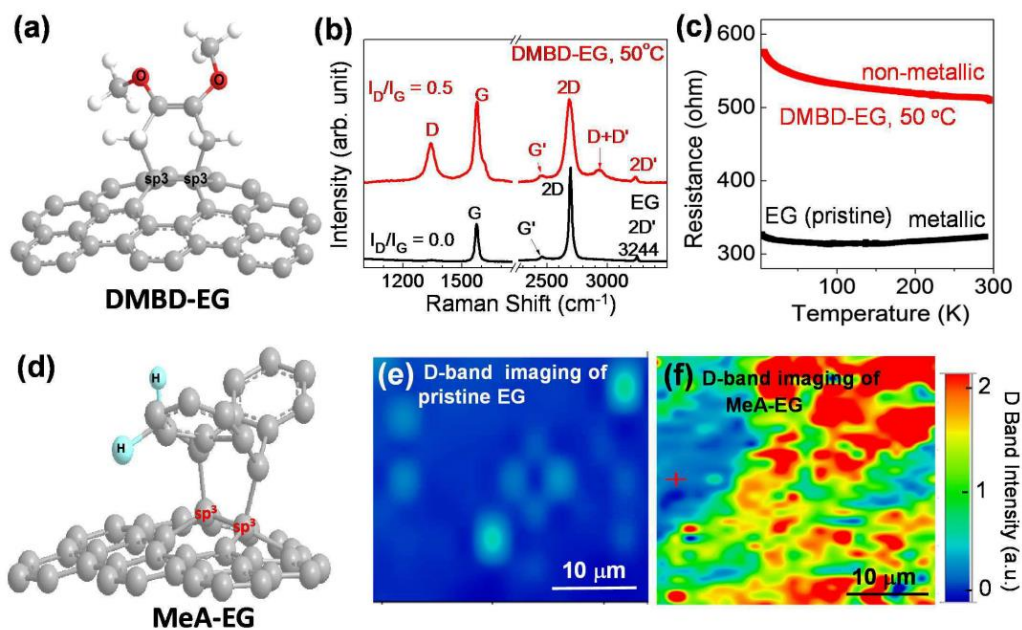

*Fig. 8* Graphene as dienophile: (a) Three-dimensional (3D) representation of 2,3-dimethoxy-1,3-butadiene (DMBD) functionalized epitaxial graphene (DMBD-EG), which shows non-planar structure due to creation of a pair of 1,2-sp$^3$ carbons in place of sp$^2$ carbons. (b) Raman spectra ($\lambda_{ex}$ = 523 nm) of pristine EG and DMBD-EG, where pristine EG shows no D-band and DMBD-EG shows $I_D/I_G$ = 0.5.[12] (c) Temperature dependence of resistance of EG, before (pristine) and after functionalization with DMBD (DMBD-EG).[12] Adapted with permission from ref[12], © 2011 American Chemical Society. (d) A 3D representation of the 9-methylanthracene (9-MeA) functionalized EG (MeA-EG), showing formation of a non-planar structure after DA chemistry. Raman D-band imaging of (e) pristine EG (before DA), and (f) after DA reaction with 9-MeA (MeA-EG), which shows moderate to high D-band intensities over the whole EG wafer.

**Future directions: a look ahead**

Experiments on the basal plane chemical functionalization of graphene have produced graphene-based materials with semiconducting and magnetic properties and thereby demonstrated the basic thesis behind our work: the possibility of using chemistry to modify the electronic and magnetic structure of graphene so as to produce a wafer patterned with dielectric, semiconducting metallic, and magnetic regions that would function as a VLSI electronic device. In the pursuit of this chemistry we have also learned that the singular electronic structure of graphene at the Dirac point can profoundly affect the course of classical pericyclic chemical processes such as the Diels-Alder reaction. There is every reason to believe that the chemistry beyond the Dirac point will prove equally fascinating and that chemistry will play a vital role in propelling graphene to assume its role as the next generation electronic material beyond silicon.

**Acknowledgements**

*The authors acknowledge financial support from DOD/DMEA under contract H94003-10-2-1003 and NSF-MRSEC through contract DMR-0820382.*